# Light pollution and the concentration of anthropogenic photons in the terrestrial atmosphere

Salvador Bará[1,*], Carmen Bao-Varela[2], and Fabio Falchi[2,3]

[1] A. Astronómica 'Ío', 15005 A Coruña, Galicia (Spain)

[2] Photonics4Life Research Group, Applied Physics Department, Universidade de Santiago de Compostela, Campus Vida, E-15782 Santiago de Compostela, Spain

[3] Istituto di Scienza e Tecnologia dell'Inquinamento Luminoso (ISTIL), 36016 Thiene, Italy

(*) Corresponding author. email: salva.bara@usc.gal
ORCID SB: https://orcid.org/0000-0003-1274-8043
ORCID CB-V: https://orcid.org/0000-0002-0602-800X
ORCID FF: https://orcid.org/0000-0002-3706-5639

**Abstract**

Light pollution can be rigorously described in terms of the volume concentration of anthropogenic photons (light quanta) in the terrestrial atmosphere. This formulation, consistent with the basic physics of the emission, scattering and absorption of light, allows one to express light pollution levels in terms of particle volume concentrations, in a completely analogous way as it is currently done with other classical pollutants, like particulate matter or molecular contaminants. In this work we provide the explicit conversion equations between the photon volume concentration and the traditional light photometry quantities. This equivalent description of the light pollution levels provides some relevant insights that help to identify artificial light at night as a standard pollutant. It also enables a complementary way of expressing artificial light exposures for environmental and public health research and regulatory purposes.

# 1. Introduction

Artificial light at night is a key factor for the improvement of the living conditions of humankind. Its production and control have provided us an essential level of freedom, enabling many activities at nighttime that otherwise would be limited by the low performance of the human visual system under typical natural night light levels. Artificial light at night is also a longtime recognized air pollutant. Already in 1979, the UN Convention on Long-range Transboundary Air Pollution, in its art. 1.a), established that

> ""Air Pollution" means the introduction by man, directly or indirectly, of substances or energy into the air resulting in deleterious effects of such a nature as to endanger human health, harm living resources and ecosystems and material property and impair or interfere with amenities and other legitimate uses of the environment, and "air pollutants" shall be construed accordingly" (United Nations, 1996) with an explicit and consistent interpretation by the UN International Law Commission in the sense that, in the context of the protection of the atmosphere,

> ""Energy" is understood to include heat, light, noise and radioactivity introduced and released into the atmosphere through human activities" (United Nations, 2018).

Artificial light at night has been shown to produce undesired disruptive effects on wildlife (Longcore and Rich, 2004; Rich and Longcore, 2006; Hölker et al., 2010; Davies et al., 2013; Svechkina et al., 2020; Gaston et al., 2021) as well as in relevant aspects of human health (Davis et al., 2001; Haim and Portnov, 2013; Stevens et al., 2014; Smolensky et al., 2015; García-Sáenz et al., 2018; Russart and Nelson, 2018; Boyce, 2022). These unwanted effects add to the progressive loss of the starry night sky (Cinzano et al., 2001; Falchi et al., 2016; Bará, 2016), whose negative consequences for the sustainability of the scientific activity of ground-based astronomical observatories (Walker, 1970; Garstang, 1989; Green et al., 2022), and the preservation of humankind's intangible cultural heritage (Marin and Jafari, 2008) have been noticed since longtime ago. Light pollution levels increase at a steady rate worldwide (Kyba et al., 2017), and their overall impact on the nocturnal environment is an issue of concern, as expressed in a growing body of recommendations (see, e.g., International Astronomical Union, 1976; Americal Medical Association, 2012; European Union, 2018; IARC, 2019; Convention on Migratory Species, 2020; International Union for Conservation of Nature, 2021; Brown et al., 2022).

Being inextricably linked to human visual experience, light has been traditionally measured and described using specific visual photometric quantities based on the *candela* ($cd$), one of the seven basic units of the international system, SI (BIPM, 2018). Visual photometric quantities are related to their physical radiometric counterparts through an internationally agreed set of well-defined conversions, mostly based on the CIE $V_\lambda$ function, the photopic spectral sensitivity of the human visual system (CIE, 1926). By applying these conversions, the spectral radiance of the electromagnetic field (usually given in energy units $W \cdot m^{-2} \cdot sr^{-1} \cdot nm^{-1}$) can be straightforwardly transformed to visual luminance ($cd \cdot m^{-2}$).

The quantum nature of light is an essential feature of our universe, and the emission, scattering and absorption processes of light are often more adequately described in terms of discrete light quanta (photons) than in terms of continuous distributions of energy (Einstein, 1905). The discrete formulation is particularly well suited for environmental studies where light plays an essential role as the basic visual input (Nilsson and Smolka, 2021; Nilsson et al, 2022), since the photochemical interactions in the eye retina take place on a photon by photon basis, and also in those cases where very low illumination levels reveal the granular nature of light. Accordingly, the spectral radiance is also commonly specified using discrete photon numbers ($photons \cdot s^{-1} \cdot m^{-2} \cdot sr^{-1} \cdot nm^{-1}$) based on the photon energy $Q_{phot} = h\nu\ joule$ ($J$), where $h = 6.626\ 070\ 15 \times 10^{-34}\ J \cdot s$ is the exact value of the Planck constant after the last SI reform (BIPM, 2018), and $\nu$ is the frequency of light ($Hz$).

The description of the environmental light exposure in terms of surface fluxes of energy or particles is of course entirely correct and, when given with enough spectral resolution and complementary information about the polarization state of the light, it provides a complete description of the light polluting field as pertinent for environmental and health sciences studies. However, expressing light pollution levels in terms of fluxes through surfaces has helped to frame light pollution as a somehow exceptional type of pollution, not clearly assimilable to the canonical air pollution forms (e.g. particulate matter of different sizes, ground-level ozone, carbon monoxide, sulfur dioxide, nitrogen dioxide, or lead), whose levels are normally expressed in terms of their volume concentrations in the environment.

As a matter of fact, light pollution levels can be equivalenty expressed in terms of the volume concentration of anthropogenic photons (in $photons \cdot m^{-3}$). This formulation gives the same quantitative results as the traditional radiometric/photometric descriptions in terms of surface flux densities and helps to highlight the fundamental similarity between light pollution and other pollution types. The quantitative relations linking the surface flux and the volume concentration formulations are explicited in this paper. Section 2 contains the main transformation equations and a brief analysis of their most relevant features. In section 3 we provide quantitative results of these transformations applied to two spectral bands of environmental interest: the human photopic ($V_\lambda$) and the Johnson-Cousins V (hereafter $V_J$). Discussion and conclusions are summarized in sections 4 and 5, respectively.

## 2. Methods

This section develops the main equations linking the surface flux and the volume concentration formulations. To that end, let us consider a small fragment $dS$ of any surface of interest, e.g. the input pupil of the eye of an individual of a given species, some elementary patch of the skin or the ground, or the entrance aperture of a light monitoring device deployed in the field.

The total radiant energy $dQ$ incident during the time $dt$ on this surface element, propagating within a small cone of directions $d\Omega$ around the direction vector $\boldsymbol{\alpha}$ and contained in the spectral wavelength interval $[\lambda, \lambda + d\lambda]$, can be written as:

$$dQ = L_\lambda(\boldsymbol{\alpha}) \cos z \; dS \; d\Omega \; d\lambda \; dt \qquad (1)$$

where $L_\lambda(\boldsymbol{\alpha})$ is the spectral radiance of the light field, generally given in energy units $W \cdot m^{-2} \cdot sr^{-1} \cdot nm^{-1}$. The direction of propagation of the radiance, $\boldsymbol{\alpha}(z, \phi)$, is usually specified in a spherical reference frame whose polar axis $Z$ is perpendicular to the surface $dS$, being $z$ the angle of propagation with respect to $Z$ and $\phi$ the azimuth, measured from any freely chosen azimuth origin. Other variables on which the radiance may depend (time, location, polarization of the light field) are not explicitly indicated in Eq.(1) for simplicity, but they should be kept in mind.

The spectral radiance $L_\lambda(\boldsymbol{\alpha})$ is the basic radiometric quantity for environmental studies, in the sense that it contains the most disaggregate information about the environmental light conditions. From this quantity any other radiometric or photometric magnitude can be immediately calculated. For instance, the total radiance $L(\boldsymbol{\alpha})$ exciting a retinal photoreceptor of sensitivity $V(\lambda)$, where $V(\lambda)$ is any generic spectral sensitivity band, coming from a source located in the direction $\boldsymbol{\alpha}$ within the eye's field of view is given by

$$L(\boldsymbol{\alpha}) = \int_{\lambda=0}^{\infty} L_\lambda(\boldsymbol{\alpha}) V(\lambda) \, d\lambda \qquad W \cdot m^{-2} \cdot sr^{-1} \qquad (2)$$

In a similar way, the irradiance $E$ within the $V(\lambda)$ band, that is, the power received per unit surface (which is the sum of the radiances arriving from all directions of the front-facing hemisphere weighted by the band sensitivity and by the cosine of the propagation angles measured from the normal to the surface) can be straightforwardly calculated as

$$E = \int_\Omega \int_{\lambda=0}^{\infty} L_\lambda(\boldsymbol{\alpha}) V(\lambda) \cos z \, d\lambda \, d\Omega \qquad W \cdot m^{-2} \qquad (3)$$

According to Eq (1) the spectral radiance is an energy density, more precisely the energy density per unit time (power), cosine-projected unit area, unit solid angle around the direction of propagation, and unit spectral interval around λ. Any radiometric or photometric quantity of environmental relevance may be easily obtained by integrating the spectral radiance with the appropriate weighting functions over the corresponding integration variables.

The radiance and irradiance, either continuous (based on $W$) or discrete (based on $photons \cdot s^{-1}$) are natively defined as flows of energy through surfaces. However, a fully equivalent and equally correct formulation of these radiometric quantities can be straightforwardly made in terms of volume concentrations of photons. To do so, let us recall that Eq.(1) can be rewritten in terms of photon numbers as

$$dQ = n_\lambda(\boldsymbol{\alpha})\, h\nu\, d\Omega\, d\lambda\, dV \qquad (4)$$

where $h\nu$ is the energy in joules per photon ($J \cdot photon^{-1}$) and $n_\lambda(\boldsymbol{\alpha})$ is the number of photons per unit volume ($V$), per unit solid angle around the propagation direction $\boldsymbol{\alpha}$ and per unit spectral interval around $\lambda$ ($n_\lambda(\boldsymbol{\alpha})$ has units $photons \cdot m^{-3} \cdot sr^{-1} \cdot nm^{-1}$). The factor $dV$ is the atmospheric volume containing the photons that will impact on the surface $dS$ propagating at the speed of light during the time $dt$ along the direction $\boldsymbol{\alpha}$, which is inclined an angle $z$ with respect to the surface normal (Fig.1). The volume of this slanted cylinder of base $dS$ is $dV = dS\, c\, dt\, \cos z$, where $c = 299\,792\,458\, m \cdot s^{-1}$ is the exact value of the speed of light (BIPM, 2018).

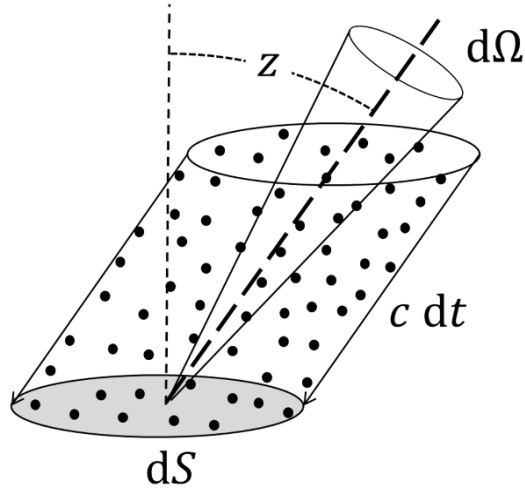

**Figure 1.** Geometry of the light propagation. The quantity $n_\lambda(\boldsymbol{\alpha})$ is the volume density of photons propagating towards the surface $dS$ per unit solid angle along the direction $\boldsymbol{\alpha}$ (at an angle $z$ with the normal to the surface) and per unit wavelength interval. The slanted cylinder volume is $dV = dS\, c\, dt\, \cos z$.

Since the energy $dQ$ is the same in Eqs.(1) and (4), and $\nu = c/\lambda$ we get

$$L_\lambda(\boldsymbol{\alpha}) = \frac{hc^2}{\lambda} n_\lambda(\boldsymbol{\alpha}) \qquad (5)$$

Equation (5) shows that the spectral radiance for each wavelength $\lambda$ is directly proportional to $n_\lambda(\boldsymbol{\alpha})$, the angular and spectral density of the volume concentration of photons.

This basic equivalence can be immediately transferred to aggregated radiometric magnitudes like the in-band radiance or the irradiance in Eqs. (2)-(3). The in-band radiance becomes:

$$L(\boldsymbol{\alpha}) = hc^2 \int_{\lambda=0}^{\infty} n_\lambda(\boldsymbol{\alpha}) \frac{V(\lambda)}{\lambda} \, d\lambda \qquad W \cdot m^{-2} \cdot sr^{-1} \qquad (6)$$

The standard practice for specifying the spectral sensitivity bands $V(\lambda)$, including the transmission functions of photometric filters and the action spectra of several physiological process, is to normalize them to 1 at its maximum. This criterion can be kept in Eq.(6) by defining a dimensionless normalized spectral sensitivity band $\Lambda(\lambda)$, derived from the original $V(\lambda)$ as

$$\Lambda(\lambda) = \frac{V(\lambda)/\lambda}{[V(\lambda)/\lambda]_{max}} \equiv \frac{V(\lambda)/\lambda}{K_0} \qquad (7)$$

where $K_0 = [V(\lambda)/\lambda]_{max}$ has dimensions of inverse length ($m^{-1}$). This allows rewriting the in-band radiance in the form

$$L(\boldsymbol{\alpha}) = hc^2 K_0 \int_{\lambda=0}^{\infty} n_\lambda(\boldsymbol{\alpha}) \, \Lambda(\lambda) \, d\lambda \qquad W \cdot m^{-2} \cdot sr^{-1} \qquad (8)$$

that can be read as

$$L(\boldsymbol{\alpha}) = hc^2 K_0 \, N(\boldsymbol{\alpha}) \qquad W \cdot m^{-2} \cdot sr^{-1} \qquad (9)$$

where $N(\boldsymbol{\alpha}) = \int_{\lambda=0}^{\infty} n_\lambda(\boldsymbol{\alpha}) \, \Lambda(\lambda) \, d\lambda$ is the volume concentration of photons of all wavelengths, weighted by $\Lambda(\lambda)$, propagating per unit solid angle in the direction $\boldsymbol{\alpha}$.

Analogously we have for the irradiance

$$E = hc^2 K_0 \int_\Omega \int_\lambda n_\lambda(\boldsymbol{\alpha}) \, \Lambda(\lambda) \, d\lambda \cos z \, d\Omega \qquad W \cdot m^{-2} \qquad (10)$$

which can be read as

$$E = hc^2 K_0 \, N \qquad (11)$$

where

$$N = \int_\Omega \int_\lambda n_\lambda(\boldsymbol{\alpha}) \, \Lambda(\lambda) \, d\lambda \cos z \, d\Omega = \int_\Omega N(\boldsymbol{\alpha}) \cos z \, d\Omega \qquad (12)$$

is the photon concentration per unit volume ($photons \cdot m^{-3}$) within the spectral band $\Lambda(\lambda)$, propagating along all directions from the front-facing hemisphere ($0 \le z \le 90°, 0 \le \phi < 360°$; $\Omega = 2\pi \, sr$) towards the element of surface on which the irradiance is measured, weighted by the cosine of the propagation angle.

## 3. Results

To get some insights about these transformations, in this section we report their specific values for the human photopic spectral sensitivity band, $V_\lambda$, and for the Johnson-Cousins $V_J$. Similar calculations can be straightforwardly made for any other band of environmental or public health interest (see section 4).

*3.1 CIE $V_\lambda$ photopic spectral sensitivity band*

The conversion between radiant units and luminous ones (i.e. radiant ones weighted by the spectral sensitivity of the human visual system and expressed in luminous units based on the $cd$) is determined by the CIE photopic spectral sensitivity function $V_\lambda$ (CIE, 1926) scaled by the constant $K_{cd}$, the SI value for the luminous efficacy of monochromatic radiation of frequency $540 \times 10^{12}$ Hz (i.e., $\lambda = 555\ nm$), defined as $K_{cd} = 683\ lm \cdot W^{-1}$ (BIPM, 2018). Within this framework the luminance $L_V(\boldsymbol{\alpha})$ of a light beam of in-band radiance $L(\boldsymbol{\alpha})$ is given by

$$L_V(\boldsymbol{\alpha}) = 683\ L(\boldsymbol{\alpha}) = 683\ hc^2 K_0\ N(\boldsymbol{\alpha}) \qquad cd \cdot m^{-2} \qquad (13)$$

and the illuminance $E_V$ corresponding to the in-band irradiance $E$ by

$$E_V = 683\ E = 683\ hc^2 K_0\ N \qquad lx \qquad (14)$$

For the CIE $V_\lambda(\lambda)$ band the function $V_\lambda(\lambda)/\lambda$ attains its maximum at $\lambda = \lambda_0 = 550.4\ nm$, resulting in $K_0 = V_\lambda(\lambda_0)/\lambda_0 = 1.8090 \times 10^6\ m^{-1}$. From these values the following conversion constants are obtained

$$\mu_V = 683\ hc^2 K_0 = 7.3581 \times 10^{-8} \quad lx\ per\ photon \cdot m^{-3} \qquad (15)$$

and

$$\eta_V = \frac{1}{683\ hc^2 K_0} = 1.3590 \times 10^7 \quad photons \cdot m^{-3}\ per\ lx \qquad (16)$$

These photon concentration conversion constants can also be expressed in any other convenient SI derived units, with less significant digits where practical or appropriate (e.g. $\eta_V \approx 13.6\ photons \cdot cm^{-3}\ per\ lx$).

The conversion constant $\mu_V$ in Eq.(15) can be interpreted relative to the normalized $\Lambda(\lambda)$ band in an analogous way as the constant $K_{cd} = 683\ lm \cdot W^{-1}$ is interpreted relative to the CIE $V_\lambda(\lambda)$ band. Recall that the latter can be rewritten as $K_{cd} = 683\ lx\ per\ W \cdot m^{-2}$. The illuminance in $lx$ is calculated by adding (integrating) the incident spectral irradiance (in $W \cdot m^{-2} \cdot nm^{-1}$) across wavelengths, weighted by the $V_\lambda(\lambda)$ function. This provides the number of band-weighted $W \cdot m^{-2}$ that, multiplied by $K_{cd}$, transforms into $lx$. In a similar way, integrating (spectrally and angularly) the photon density $n_\lambda(\boldsymbol{\alpha})$ in Eq.(12) weighted by the $\Lambda(\lambda)$ function provides the number of band-weighted $photons \cdot m^{-3}$, $N$, that after multiplication by $\mu_V$, transforms into $lx$. It is important to keep in mind that $N$ is a band-weighted volume density: this means that two monochromatic radiations of wavelengths $\lambda_1$, $\lambda_2$, will produce an equal amount of illuminance (or of any other photometric luminous quantity) if their respective basic photon densities $n_\lambda(\boldsymbol{\alpha})$ in Eq.(4) are in the ratio $n_{\lambda_1}(\boldsymbol{\alpha})/n_{\lambda_2}(\boldsymbol{\alpha}) = \Lambda(\lambda_2)/\Lambda(\lambda_1)$. This echoes the well-known fact that the same amount of illuminance can be obtained with two monochromatic radiations whose spectral irradiances are in the ratio $E(\lambda_1)/E(\lambda_2) = V_\lambda(\lambda_2)/V_\lambda(\lambda_1)$.

The in-band, cosine weighted photon volume concentration $\eta_V$ corresponding to 1 $lx$, Eq.(16), is not as small as at a first glance these figures could suggest. This volume concentration is equivalent to an in-band photon surface flux density of $\eta_V \times c = 4.0743 \times 10^{15}\ photons \cdot s^{-1} \cdot m^{-2}\ per\ lux$, consistent with the expected value. Recall that photons just need about 3.3 ns time to travel 1 m. Given a 1m x 1m detector surface, photons span in that time a volume of 1 m³ that contains $1.3590 \times 10^7$ in-band, cosine weighted photons per lux incident on that surface.

### 3.2 Johnson-Cousins $V_J$ band

Another spectral band widely used in environmental light pollution monitoring and research is the Johnson-Cousins *V* (Bessell, 1999; Bessell and Murphy, 2012), here denoted $V_J(\lambda)$. This band is located in the central region of the optical spectrum, has a bandwidth of 90.89 nm and peaks at 529 nm with a normalized maximum value equal to 1. The Johnson-Cousins $V_J$ band has been extensively used as standard for reporting the anthropogenic brightness of the night sky, and several key measuring devices (Hänel et al., 2018) and iconic light pollution quantitative models (Falchi et al., 2016) are designed and built based on it.

The photon volume concentration corresponding to the irradiance in this band can be easily calculated from Equation (11), taking into account that in this case the function $V_J(\lambda)/\lambda$ peaks at $\lambda_0 = 526.5\ nm$, resulting in $K_0 = V_J(\lambda_0)/\lambda_0 = 1.8959 \times 10^6\ m^{-1}$. This immediately leads to the conversion factors

$$\mu_J = hc^2 K_0 = 0.1129 \times 10^{-9}\quad W \cdot m^{-2}\ per\ photon \cdot m^{-3} \qquad (17)$$

$$\eta_J = \frac{1}{hc^2 K_0} = 8.8568 \times 10^9 \quad photons \cdot m^{-3}\ per\ W \cdot m^{-2} \qquad (18)$$

As an example of the practical significance of this conversion factors, Figure 2 shows the photon volume concentrations $N$ in $log_{10}\ (photons \cdot m^{-3})$ associated with the ground level horizontal irradiance produced by the artificial brightness of the night sky, in an area comprising the Iberian Peninsula, Northern Maghreb (Northern Algeria, Morocco, and Tunisia), Southern France, and Balearic Islands. The map, in log decimal grayscale, shows the large increase of the volume concentration of anthropogenic photons in and around metropolitan areas, as well as its pervasive presence in wide inhabited regions, including the coastal waters of the Atlantic Ocean and the Mediterranean sea. The anthropogenic sky irradiance has been computed by following the procedures and methods described in Falchi and Bará (2021).

Note that in addition to the photon concentrations produced by the artificial light scattered in the atmosphere, shown in Fig. 2, there is also in many places (urban regions, areas surrounding roadways, industrial installations,...) a noticeable contribution of direct radiance from the nearby streetlights, whose associated photon volume concentrations would add to the ones shown in this map.

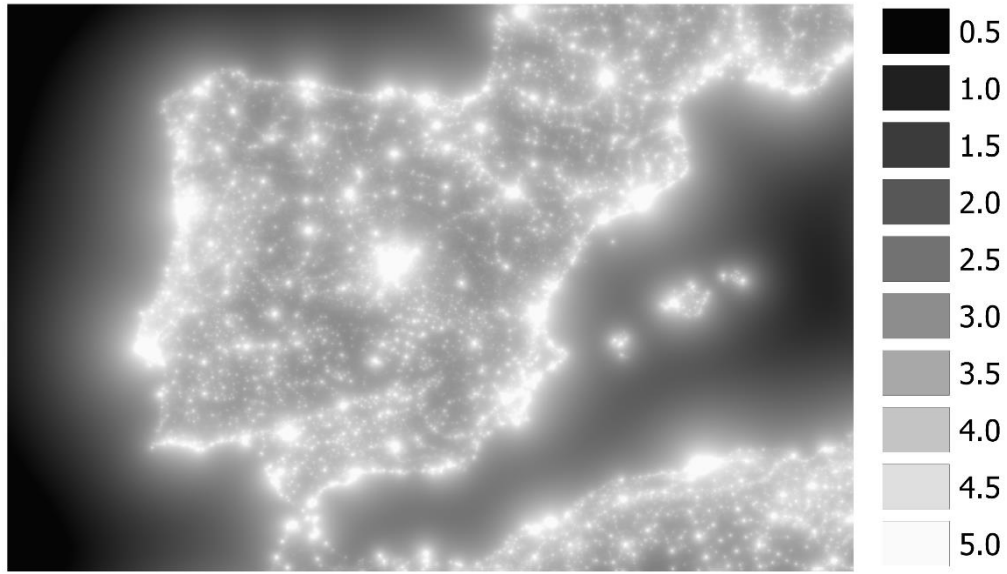

**Figure 2.** Ground-level photon volume concentrations ($N$) in the Johnson-Cousins $V_J$ band, associated with the horizontal irradiance produced by the artificial brightness of the night sky, in an area comprising the Iberian Peninsula, Northern Maghreb (Northern Algeria, Morocco, and Tunisia), Southern France, and Balearic Islands. CRS: EPSG:25830 - ETRS89 / UTM zone 30N. Pixel width 409.44 m. Grayscale in $log_{10}\ (photons \cdot m^{-3})$.

## 4. Discussion

In the above sections the correspondence between the usual radiometric and photometric quantities expressed as surface flux densities and their associated photon volume concentrations has been explicited in a general way (section 2) and particularized for the spectral sensitivity of the human visual system and the Johnson-Cousins $V_J$ band (section 3). The same approach can be applied to any other spectral sensitivity band of environmental or public health interest, as e.g. different types of opsins (Govardovskii et al., 2000), wildlife action spectra (van Grunsven et al., 2014; Donners et al, 2018; Longcore et al., 2018; Nilsson and Smolka, 2021), or the α-opic sensitivities of human retinal photoreceptors mediating non-image forming effects of light, relevant for hormone regulation and physiological rhythms synchronization, among others (CIE, 2018; Schlangen and Price, 2021).

The conceptualization of light pollution as an increased volume concentration of polluting particles in the nocturnal environment above the expected natural values, first presented and developed in two seminal papers by Cinzano and Falchi (2012, 2014), can be easily extended to any of the above quoted bands by using Eqs. (8) and (10) of this paper. Reference values for the natural baseline in diverse bands may be found in Masana et al. (2021, 2022).

The proposed description does not rely on a unique total volume density of photons, but on the volume density of the subset of photons relevant for the photometric magnitude under study, which in general will be different for each magnitude. In case of the spectral radiance in a given direction ($\alpha$), the relevant photon volume density is $n_\lambda(\alpha)$, Eq.(4), i.e. that of the photons propagating per unit solid angle and unit spectral interval along the prescribed direction. In case of the total radiance/luminance in that direction, the relevant photon volume density is $N(\alpha)$, defined in Eq.(9) after an in-band spectral integration. In case of the total irradiance/illuminance it is the volume density $N$ of the photons propagating in all directions (cosine and band-weighted) towards the chosen elementary surface, defined in Eq.(12). If the radiometric magnitude under study were the total radiant energy within a volume, the relevant density would be related to the total number of photons per unit volume. Other metrics may require their own density definitions (see, e.g. Jechow and Hölker, 2019). The main takeaway message is that every radiometric/photometric magnitude can be equivalently expressed in terms of the volume density of the appropriate subset of photons. Specifying volume densities of different subsets of photons (classified by their properties) is analogous to specifying volume densities of different subsets of particulate matter (classified by their properties, e.g. sizes as in PM2.5, PM10, etc.). The concentration of PM2.5 is not the total concentration of material particles in the air, but the concentration of the particles of a given size range, relevant for a definite purpose; the same approach is followed here when defining the photon concentrations described in this paper.

The formulation of environmental light exposures in terms of volume concentrations of photons is a complement, rather than a substitute for the traditional specification of lighting levels in terms of radiometric and photometric surface fluxes. However, future light pollution regulations may take advantage of this approach to specify exposure limits (also) in terms of maximum allowed atmospheric concentrations of anthropogenic light particles, seamlessly unifying them with the current regulations of other well-known atmospheric pollutants.

## 5. Conclusions

The light pollution levels in the nocturnal environment can be rigorously expressed in terms of the volume concentration of anthropogenic light particles. The basic equations of this formulation and their application to any spectral sensitivity band, including visual and non-visual ones of environmental and public health relevance, are described in this work. This equivalent formulation of the light pollution exposures, consistent with the basic physical processes of emission, scattering, and absorption of light, provides a unified framework to understand and manage artificial light at night as a conventional air pollutant.

**Acknowledgements**

Any scientific research work benefits from multiple exchanges with fellow scientists, and this is no exception. Special thanks are due, among others, to Raul Lima and Martin Pawley for insightful comments and criticisms on light pollution issues. Anonymous reviewers provided very useful suggestions that helped to improve this paper. Any remaining error of this paper is of course the sole responsibility of the authors.

**Funding sources**

CB acknowledges funding from Xunta de Galicia/FEDER, grant ED431B 2020/29.

**References**

American Medical Association, Ama, 2012. Light Pollution: adverse health effects of nighttime lighting. Chicago, Illinois (USA. In: Proceedings of the American MedicalAssociation House of Delegates, 161st Annual Meeting, pp. 265–279. Online: https://www.ama-assn.org/sites/ama-assn.org/files/corp/media-browser/public/hod/a12-csaph-reports_0.pdf.

Bará, S., 2016. Anthropogenic disruption of the night sky darkness in urban and rural areas. R. Soc. Open Sci. 3, 160541 https://doi.org/10.1098/rsos.160541.

Bessell, M.S., 1990. UBVRI Passbands, vol. 102. Publications of the Astronomical Society of the Pacific, pp. 1181–1199. https://doi.org/10.1086/132749.

Bessell, M., Murphy, S., 2012. Spectrophotometric libraries, revised photonic passbands, and zero points for UBVRI, hipparcos, and tycho photometry. Publ. Astron. Soc. Pac. 124, 140–157.

BIPM, 2018. Resolution 1 of the 26th CGPM: on the revision of the international system of units (SI). https://www.bipm.org/en/committees/cg/cgpm/26-2018/resolution-1. (Accessed 11 April 2022).

Boyce, P., 2022. Light, lighting and human health. Light. Res. Technol. 54 (2), 101–144. https://doi.org/10.1177/14771535211010267.

Brown, T.M., Brainard, G.C., Cajochen, C., Czeisler, C.A., Hanifin, J.P., Lockley, S.W., et al., 2022. Recommendations for daytime, evening, and nighttime indoor light exposure to best support physiology, sleep, and wakefulness in healthy adults. PLoS Biol. 20 (3), e3001571 https://doi.org/10.1371/journal.pbio.3001571.

Cayrel, R., 1979. Identification and Protection of Existing and Potential Observatory Sites. Trans. Int. Astron. Union 17 (1), 215–223. https://doi.org/10.1017/S0251107X00010798.

CIE, 1926. Commission Internationale de l'Eclairage Proceedings, 1924. Cambridge University Press, Cambridge.

CIE, 2018. Commission Internationale de l'Eclairage. CIE System for Metrology of Optical Radiation for ipRGC-Influenced Responses to Light. Publication CIE S 026/E. https://doi.org/10.25039/S026.2018.

Cinzano, P., Falchi, F., 2012. The propagation of light pollution in the atmosphere. Mon. Not. Roy. Astron. Soc. 427, 3337–3357. https://doi.org/10.1111/j.1365-2966.2012.21884.x.

Cinzano, P., Falchi, F., 2014. Quantifying light pollution. J. Quant. Spectrosc. Radiat. Transf. 139, 13–20. https://doi.org/10.1016/j.jqsrt.2013.11.020.

Cinzano, P., Falchi, F., Elvidge, C., 2001. The first world atlas of the artificial night sky brightness. Mon. Not. Roy. Astron. Soc. 328, 689–707. https://doi.org/10.1046/j.1365-8711.2001.04882.x.

Convention on the Conservation of Migratory Species of Wild Animals, 2020. Light pollution guidelines for wildlife. Thirteenth meeting of the conference of the parties to CMS , 15.02.2020. India, CMS Resolution 13.5. Publish date 08 April 2020 1–2. https://www.cms.int/sites/default/files/document/cms_cop13_res.13.5_light-pollution-guidelines_e.pdf. (Accessed 5 August 2022).

Davies, T.W., Bennie, J., Inger, R., Gaston, K.J., 2013. Artificial light alters natural regimes of night-time sky brightness. Sci. Rep. 3, 1722. https://doi.org/10.1038/srep01722.

Davis, S., Mirick, D.K., Stevens, R.G., 2001. Night shift work, light at night, and risk of breast cancer. J. Natl. Cancer Inst. 93, 1557–1562. https://doi.org/10.1093/jnci/93.20.1557.

Donners, M., van Grunsven, R.H.A., Groenendijk, D., van Langevelde, F., Bikker, J.W., Longcore, T., Veenendaal, E.M., 2018. Colors of attraction: modeling insect flight to light behavior. J. Exp. Zool. 329, 434–440. https://doi.org/10.1002/jez.2188.

Einstein, A., 1905. Über einen die Erzeugung und Verwandlung des Lichtes betreffenden heuristischen Gesichtspunkt [On a Heuristic Point of View about the Creation and Conversion of Light]. Ann. Phys. 17 (6), 132–148 (in German).

European Union, 2018. EU Green Public Procurement criteria for road lighting andtraffic signals. http://ec.europa.eu/environment/gpp/pdf/toolkit/181210_EU_GPP_criteria_road_lighting.pdf.

Falchi, F., Bará, S., 2021. Computing light pollution indicators for environmental assessment. Nat Sci. e10019. https://doi.org/10.1002/ntls.10019.

Falchi, F., Cinzano, P., Duriscoe, D., Kyba, C.C.M., Elvidge, C.D., Baugh, K., Portnov, B. A., Rybnikova, N.A., Furgoni, R., 2016. The new world atlas of artificial night sky brightness. Sci. Adv. 2, e1600377 https://doi.org/10.1126/sciadv.1600377.

García-Sáenz, A., de Miguel, A.S., Espinosa, A., Valentin, A., Aragones, N., Llorca, J., Amiano, P., Sanchez, V.M., Guevara, M., Capelo, R., et al., 2018. Evaluating the association between artificial light-at-night exposure and breast and prostate cancer risk in Spain (MCC-Spain study). Environ. Health Perspect. 126 (4), 047011 https://doi.org/10.1289/EHP1837.

Garstang, R.H., 1989. Night-sky Brightness at Observatories and Sites, vol. 101. Publications of the Astronomical Society of the Pacific, pp. 306–329. https://doi.org/10.1086/132436.

Gaston, K.J., Ackermann, S., Bennie, J., Cox, D.T.C., Phillips, B.B., Sánchez de Miguel, A., Sanders, D., 2021. Pervasiveness of biological impacts of artificial light at night. Integr. Comp. Biol. 61 (3), 1098–1110. https://doi.org/10.1093/icb/icab145.

Govardovskii, V.I., Fyhrquist, N., Reuter, T., Kuzmin, D.G., Donner, K., 2000. In search of the visual pigment template. Vis. Neurosci. 17 (4), 509–528. https://doi.org/10.1017/s0952523800174036.

Green, R.F., Luginbuhl, C.B., Wainscoat, R.J., et al., 2022. The growing threat of light pollution to ground-based observatories. Astron. AstroPhys. Rev. 30, 1. https://doi.org/10.1007/s00159-021-00138-3.

Haim, A., Portnov, B., 2013. Light Pollution as a New Risk Factor for Human Breast and Prostate Cancers. Springer, Heidelberg. https://doi.org/10.1007/978-94-007-6220-6.

Hänel, A., Posch, T., Ribas, S.J., Aubé, M., Duriscoe, D., Jechow, A., Kollath, Z., Lolkema, D.E., Moore, C., Schmidt, N., Spoelstra, H., Wuchterl, G., Kyba, C.C.M., 2018. Measuring night sky brightness: methods and challenges. J. Quant. Spectrosc. Radiat. Transfer 205, 278–290. https://doi.org/10.1016/j.jqsrt.2017.09.008.

Hölker, F., Wolter, C., Perkin, E.K., Tockner, K., 2010. Light pollution as a biodiversity threat. Trends Ecol. Evol. 25, 681–682. https://doi.org/10.1016/j.tree.2010.09.007.

IARC Monographs Vol 124 group, Ward, E.M., Germolec, D., Kogevinas, M., et al., 2019.Carcinogenicity of night shift work. Lancet Oncol. 20 (8), 1058–1059. https://doi.org/10.1016/S1470-2045(19)30455-3.

International Union for Conservation of Nature, IUCN, 2021. Taking Action to ReduceLight Pollution. IUCN World Conservation Congress. Resolution 084. https://www.iucncongress2020.org/motion/084.

Jechow, A., Hölker, F., 2019. How dark is a river? Artificial light at night in aquatic systems and the need for comprehensive night-time light measurements. Wiley Interdisciplinary Reviews: Water 6 (6), e1388. https://doi.org/10.1002/wat2.1388.

Kyba, C.C.M., Kuester, T., Sánchez de Miguel, A., Baugh, K., Jechow, A., Hölker, F., Bennie, J., Elvidge, C.D., Gaston, K.J., Guanter, L., 2017. Artificially lit surface of Earth at night increasing in radiance and extent. Sci. Adv. 3, e1701528 https://doi.org/10.1126/sciadv.1701528.

Longcore, T., Rich, C., 2004. Ecological light pollution. Front. Ecol. Environ. 2, 191–198. https://doi.org/10.1890/1540-9295(2004)002[0191:ELP]2.0.CO;2.

Longcore, T., Rodríguez, A., Witherington, B., Penniman, J.F., Herf, L., Herf, M., 2018. Rapid assessment of lamp spectrum to quantify ecological effects of light at night. J. Exp. Zool. 329, 511–521. https://doi.org/10.1002/jez.2184.

Marín, C., Jafari, J., 2008. StarLight: A Common Heritage; StarLight Initiative La Palma Biosphere Reserve. Instituto De Astrofísica De Canarias, Government of The Canary Islands, Spanish Ministry of The Environment, UNESCO-MaB, Canary Islands, Spain.

Masana, E., Carrasco, J.M., Bará, S., Ribas, S.J., 2021. A multi-band map of the natural night sky brightness including Gaia and Hipparcos integrated starlight. Monthly Notices of the Royal Astronomical Society 501, 5443–5456. doi: 10.1093/mnras/staa4005

Masana, E., Bará, S., Carrasco, J.M., Ribas, S.J., 2022. An enhanced version of the Gaia map of the brightness of the natural sky. International Journal of Sustainable Lighting 24 (1), 1–12. https://doi.org/10.26607/ijsl.v24i1.119.

Nilsson, D.-E., Smolka, J., 2021. Quantifying biologically essential aspects ofenvironmental light. J. R. Soc. Interface 18, 20210184. https://doi.org/10.1098/rsif.2021.0184.

Nilsson, D.-E., Smolka, J., Bok, M., 2022. The vertical light-gradient and its potentialimpact on animal distribution and behavior. Front. Ecol. Evol. 10, 951328 https://doi.org/10.3389/fevo.2022.951328.

Rich, C., Longcore, T. (Eds.), 2006. Ecological Consequences of Artificial Night Lighting. Island Press, Washington, D.C.

Russart, K.L.G., Nelson, R.J., 2018. Light at night as an environmental endocrine disruptor. Physiol. Behav. 190, 82–89. https://doi.org/10.1016/j.physbeh.2017.08.029.

Schlangen, L.J.M., Price, L.L.A., 2021. The lighting environment, its metrology, and nonvisual responses. Front. Neurol. 12, 624861 https://doi.org/10.3389/fneur.2021.624861.

Smolensky, M.H., Sackett-Lundeen, L.L., Portaluppi, F., 2015. Nocturnal light pollution and underexposure to daytime sunlight: complementary mechanisms of circadian disruption and related diseases. Chronobiol. Int. 32 (8), 1029–1048. https://doi.org/10.3109/07420528.2015.1072002.

Stevens, R.G., Brainard, G.C., Blask, D.E., Lockley, S.W., Motta, M.E., 2014. Breast cancer and circadian disruption from electric lighting in the modern world. CA A Cancer J. Clin. 64, 207–218. https://doi.org/10.3322/caac.21218.

Svechkina, A., Portnov, B.A., Trop, T., 2020. The impact of artificial light at night on human and ecosystem health: a systematic literature review. Landsc. Ecol. 35, 1725–1742. https://doi.org/10.1007/s10980-020-01053-1.

United Nations, 1996. 1979 Convention on Long-Range Transboundary Air Pollution and its Protocols [E/]ECE/EB.AIR/50. UN, New York ; Geneva (last accessed May 10th, 2022). https://digitallibrary.un.org/record/237573?ln=en.

United Nations, 2018. Report of the International Law Commission. Seventieth Session (30 April 1 June and 2 July 10 August 2018) General Assembly, Official Records,Seventy Third Session. UN, p. 171. Supplement No. 10 (A/73/10), VI. Protection of the atmosphere. https://legal.un.org/ilc/reports/2018/. (Accessed 10 May 2022).

van Grunsven, R.H.A., Donners, M., Boekee, K., Tichelaar, I., van Geffen, K.G., Groenendijk, D., Berendse, F., Veenendaal, E.M., 2014. Spectral composition of light sources and insect phototaxis, with an evaluation of existing spectral response models. J. Insect Conserv. 18 (2), 225–231. https://doi.org/10.1007/s10841-014-9633-9.

Walker, M.F., 1970. The California site survey. Publ. Astron. Soc. Pac. 82, 672–698. https://www.jstor.org/stable/40674892.